# A quantitative and qualitative open citation analysis of retracted articles in the humanities

Ivan Heibi – ivan.heibi2@unibo.it – https://orcid.org/0000-0001-5366-5194

Research Centre for Open Scholarly Metadata, Department of Classical Philology and Italian Studies, University of Bologna, Bologna, Italy

Digital Humanities Advanced Research Centre (/DH.arc), Department of Classical Philology and Italian Studies, University of Bologna, Bologna, Italy

Silvio Peroni – silvio.peroni@unibo.it – https://orcid.org/0000-0003-0530-4305

Research Centre for Open Scholarly Metadata, Department of Classical Philology and Italian Studies, University of Bologna, Bologna, Italy

Digital Humanities Advanced Research Centre (/DH.arc), Department of Classical Philology and Italian Studies, University of Bologna, Bologna, Italy

## Abstract

In this article, we show and discuss the results of a quantitative and qualitative analysis of open citations to retracted publications in the humanities domain. Our study was conducted by selecting retracted papers in the humanities domain and marking their main characteristics (e.g., retraction reason). Then, we gathered the citing entities and annotated their basic metadata (e.g., title, venue, subject, etc.) and the characteristics of their in-text citations (e.g., intent, sentiment, etc.). Using these data, we performed a quantitative and qualitative study of retractions in the humanities, presenting descriptive statistics and a topic modeling analysis of the citing entities' abstracts and the in-text citation contexts. As part of our main findings, we noticed that there was no drop in the overall number of citations after the year of retraction, with few entities which have either mentioned the retraction or expressed a negative sentiment toward the cited publication. In addition, on several occasions, we noticed a higher concern/awareness when it was about citing a retracted publication, by the citing entities belonging to the health sciences domain, if compared to the humanities and the social science domains. Philosophy, arts, and history are the humanities areas that showed the higher concern toward the retraction.

# 1. Introduction

Retraction is a way to correct the literature and alerting readers on erroneous materials in the published literature. A retraction should be formally accompanied by a retraction notice – a document that justify such a retraction. Reasons for retraction includes plagiarism, peer review manipulation, unethical research, etc. (Barbour et al., 2009)

Several works in the past studied and uncovered important aspects regarding this phenomenon, such as the reasons for retraction (Casadevall et al., 2014; Corbyn, 2012), the temporal characteristics of the retracted articles (Bar-Ilan & Halevi, 2018), their authors' countries of origin (Ataie-Ashtiani, 2018), and the impact factor of the journals publishing them (Campos-Varela et al., 2020) (Fang & Casadevall, 2011). Other works have analyzed authors with a higher number of retractions (Brainard, 2018), and the scientific impact, technological impact, funding impact, and Altmetric impact in retractions (Feng et al., 2020). Other studies focused on the retraction in the medical and biomedical domain (Gaudino et al., 2021; Campos-Varela et al., 2020; Gasparyan et al., 2014).

Scientometricians have also proposed several works on retraction based on quantitative data. For instance, several works (Lu et al., 2013; Azoulay et al., 2017; Mongeon & Larivière, 2016; Shuai et al., 2017) focused on showing how a single retraction could trigger citation losses through an author's prior body of work. Bordignon (2020) investigated the different impacts that negative citations in articles and comments posted on post-publication peer review platforms have on the correction of science, while Dinh et al. (2019) applied descriptive statistics and ego-network methods to examine 4,871 retracted articles and their citations before and after retraction. Other authors focused on the analysis of the citations made before the retraction (Bolland et al., 2021) and on a specific reason of retraction such as misconduct (Candal-Pedreira et al., 2020). The studies that considered only one retraction case usually observed also the in-text citations and the related citation context in the articles citing retracted publications (van der Vet & Nijveen, 2016; Bornemann-Cimenti et al., 2016; Luwel et al., 2019; Schneider et al., 2020).

Although a citation analysis toward the retraction has been done several times in STEM (Science, Technology, Engineering, and Mathematics) disciplines, less attention has been given to the humanities domain. One of this rare analysis done in the humanities domain has been recently presented by Halevi (2020), which considered two examples of retracted articles and showed their continuous post-retraction citations.

Our study wants to expand the works concerning the analysis of citations to retracted publications in the humanities domain. By combining quantitative analysis, with a quantification of citations and their related characteristics/metadata, and qualitative analysis, through a subjective examination of aspects related to the quality of the citations (e.g., the reason of a citation based on the examination/interpretation of its in-text citation context), we aimed at understanding this phenomenon in the humanities, which gained little attention in the past literature. In particular, the research questions (RQ1-RQ3) we aimed to address are:

- RQ1: How did scholarly research cite retracted humanities publications before and after their retraction?
- RQ2: Did all the humanities areas behave similarly concerning the retraction phenomenon?
- RQ3: What were the main differences in citing retracted publications between STEM disciplines and the Humanities?

In this paper, we used a methodology developed to gather, characterize, and analyze incoming citations of retracted publications (Heibi & Peroni, 2022), and we adapted it for the case of the humanities[1]. The citation analysis is based on collections of open citations, i.e., data are structured, separate, open, identifiable, available (Peroni & Shotton, 2018, 2020).

# 2. Data gathering

The workflow followed to gather and analyze the data in this study is based on the methodology introduced in (Heibi & Peroni, 2022), briefly summarized in Figure 1. The first two phases of the methodology are dedicated to the collection and characterization of the entities that have cited the retracted publications. The third phase is focused in analyzing the information annotated in the first two phases to summarize quantitatively the data collected. The fourth and final phase, applies a topic modeling analysis (Barde & Bainwad, 2017) on the textual information (extracted from the full text of the citing entities) and builds a set of dynamic visualizations to enable an overview and investigation of the generated topics. The data gathering of our study is detailed in the following sections.

---

[1] It is worth mentioning that we did not describe the methodology adopted in full here due to space constraints.

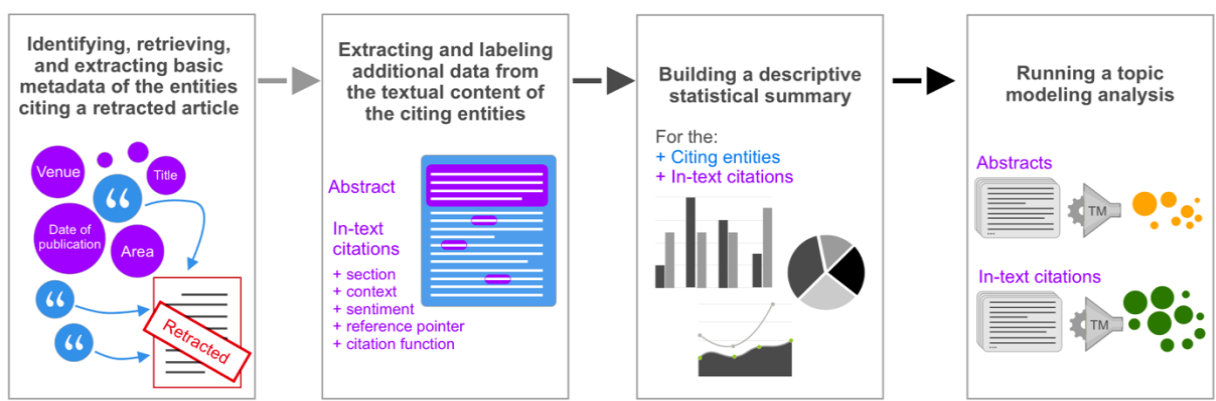

**Figure 1.** A summarizing schema representing the methodology in its four phases: (1) identifying, retrieving, and characterizing the citing entities, (2) extracting and labeling additional features based on the citing entities contents, (3) building a descriptive statistical summary, and (4) running a topic modeling analysis

## 2.1. Retraction in the humanities

First, we wanted to have a descriptive statistical overview of the retractions in the humanities as a function of crucial features (e.g., reasons of retraction) to help us define the set of retractions to use as input in the next phases. Thus, we queried the Retraction Watch database (http://retractiondatabase.org) (Collier, 2011) searching for all the retracted publications labelled as humanities (marked with "HUM" in the database). Thus, the humanities domain considered in this work is based on the subject classification used by Retraction Watch, i.e., the subjects under the macro category "(HUM) Humanities". Then we have classified the results as a function of three parameters: (a) the year of the retraction, (b) the subject area of the retracted publications (architecture, arts, etc.), and (c) the reason(s) for the retraction. We collected an overall number of 474 publications, the earlier retraction occurred in 2002, while the last year of retraction we obtained was 2020.

As shown in Figure 2, we noticed an increasing trend throughout the years, with some exception, in particular, we observed that the highest number of retractions per year was 119 in 2010, probably due to an investigation and a massive retraction of several articles belonging to one author, i.e., Joachim Boldt (Brainard, 2018). When looking at the subject areas, we noticed that most of the retractions are related to *arts* and *history*, while plagiarism motives[2] were by far the most representative ones, confirming the observation in (Halevi, 2020). Most of the retracted publications (88%) are of article type (i.e., labeled in Retraction Watch as either "Conference Abstract/Paper", "Research Article", or "Review Articles"). Book chapters/References represent 8% of the total, the rest are "Commentary/Editorials" (1%), and other residual types (3%, e.g., letters, case reports, articles in press, etc.).

---

[2] A complete list of reasons accompanied by a description is provided by Retraction Watch at https://retractionwatch.com/retraction-watch-database-user-guide/retraction-watch-database-user-guide-appendix-b-reasons/.

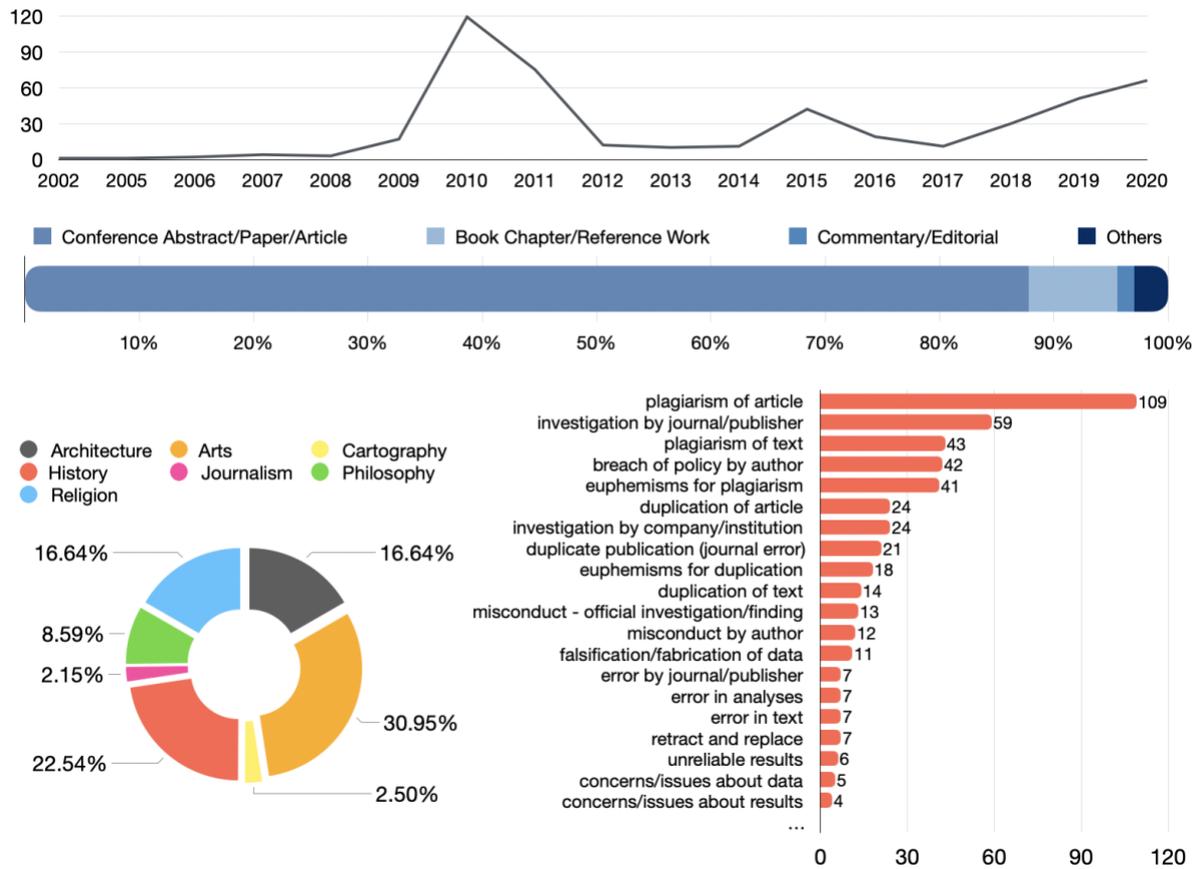

**Figure 2.** Retractions in the humanities domain with respect to three different features: the year of retraction (line chart), the subject areas of the retracted publications (ring chart), the type of the retracted publication (large horizontal bar), and the reasons for retraction (horizontal bar chart). Based on the data retrieved from the Retraction Watch database in June 2021.

## 2.2. Retracted publications set and their citations

Since the focus of our study is on the analysis of citations to fully retracted publications, we excluded all the retracted publications collected in the previous step that did not receive at least one citation according to two open citation databases: Microsoft Academic Graph (MAG, https://www.microsoft.com/en-us/research/project/microsoft-academic-graph/) (Wang et al., 2020) and OpenCitations' COCI (https://opencitations.net/index/coci) (Heibi et al., 2019). MAG is a knowledge graph that contains the scientific publication records, citations, authors, institutions, journals, conferences, and fields of study. It also provides a free REST API service to search, filter, and retrieve its data. COCI is a citation index which contains details of all the DOI-to-DOI citation links retrieved by processing the open bibliographic references available in Crossref (Hendricks et al., 2020), and it can be queried using open and free REST APIs. We decided to not use other proprietary and non-open databases since we aimed at making our workflow and our results as reproducible as possible.

After querying COCI and MAG[3], we found that 85 retracted items (out of 474) had at least one citation (2054 citations). We manually checked the dataset from possible mistakes introduced by the collections. Indeed, some of the citing entities identified in MAG either did not include a bibliographic reference to any of the retracted publications, or the retracted publication in consideration was not cited in the content of the citing entity (although present in its reference list), or the citing entities' type did not refer to a scholarly publication (e.g., bibliography, retraction notice, presentation, data repository). There was also one retracted article for duplication "The Nature of Creativity" by Sternberg (2006) that received 1,050 citations. This retracted article contains a substantial amount of content previously published by the same author on several of his previous works and it was the fourth retracted article of the same author who used to cite himself at a high rate and not doing enough to encourage diversity in psychology research. We decided to exclude it from our study to reduce bias in the results. Following these considerations, the final number of retracted publications considered was 84, involving a total number of 935 unique citing entities. As shown in the bubble chart in Figure 3, most of the citing entities (i.e., 891) were included in MAG, 388 were included in COCI, and they shared 344 entities.

---

[3] We used their REST APIs in June 2021 for retrieving citation information.

Although the retracted items identified so far were all in the humanities domain according to the categories specified in Retraction Watch, an item might have other non-humanities subjects associated with it. Sometimes, these non-humanities subjects might be more representative of the content of the retracted document and, thus, they might generate an unwanted bias for the rest of the analysis. For instance, consider the retracted article *"The Good, the Bad, and the Ugly: Should We Completely Banish Human Albumin from Our Intensive Care Units?"* (Boldt, 2000). In Retraction Watch, the subjects associated with it were medicine and journalism. Yet, when we checked the full-text of the article, we noticed that argumentations close to journalism are very few and, as such, the article should not be considered as belonging to humanities research.

To avoid considering these peculiar publications in our analysis, we devised a mechanism to help us evaluating the affinity of each retracted item to the humanities domain. We assigned to each retracted item in the list (84) an initial score of 1, named *hum_affinity* – this value ranges from 0 (i.e. very low) to 5 (i.e. very high). The final value of *hum_affinity* for each retracted item is calculated as follows:

1. We assigned to each retracted item additional subject categories obtained by searching the venue where it was published in external databases – we used Scimago classification (https://www.scimagojr.com/) for journals and the Library of Congress Classification (LCC, https://www.loc.gov/catdir/cpso/lcco/) for books/book chapters.
2. If both the Retraction Watch subjects and those gathered in step (1) included at least one subject identifying a discipline in the humanities, we added 1 to *hum_affinity* of that item.

3. If all the Retraction Watch subjects are part of the humanities domain, we added another 1 to *hum_affinity* of that item.

4. If the title of the retracted item has a clear affinity to the humanities (e.g. "The Origins of Probabilism in Late Scholastic Moral Thought"), we added another 1 to *hum_affinity* of that item.

5. Finally, we provided a subjective score of -1, 0, or 1 based on the abstract of the item. For instance, we assigned 1 to the abstract of the retracted article of Mößner (2011): "… This paper aims at a more thorough comparison between Ludwik Fleck's concept of thought style and Thomas Kuhn's concept of paradigm. Although some philosophers suggest that these two concepts …".

The pie chart in Figure 3 shows how we classified the retracted publications and those citing them according to their *hum_affinity* score. To narrow our analysis and reduce bias, we decided to consider only the retracted publications (and their corresponding citing entities) having a medium or high *hum_affinity* score (i.e., ≥ 2). A total of 12 retracted publications have been excluded from the analysis (i.e., *hum_affinity*<2) along with their 257 citations. A list of the excluded retracted publications is available at the zenodo repository (Heibi & Peroni, 2021b). At the end of this phase, the final number of retracted items we considered was 72, with a total of 678 citing entities.

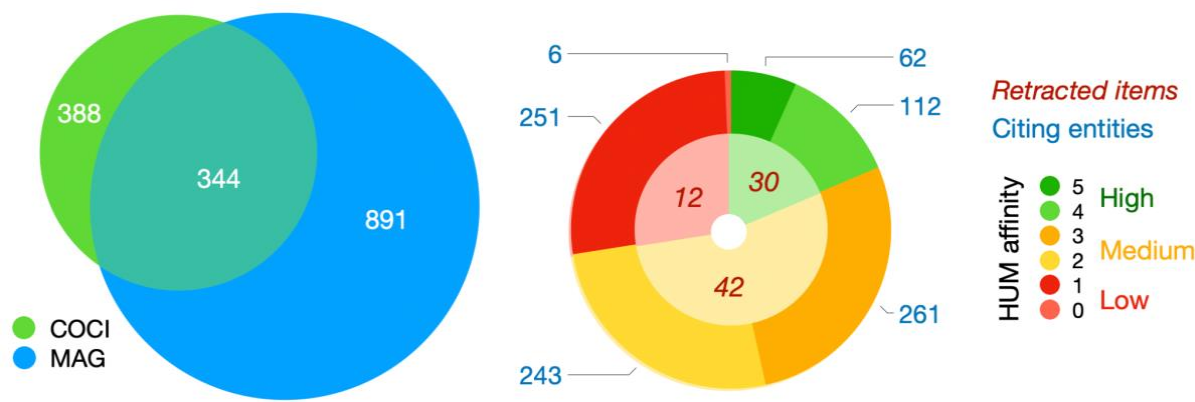

**Figure 3.** A Venn diagram (bubble chart) to plot the number of entities gathered from MAG (Microsoft Academic Graph) and COCI (OpenCitations Index of Crossref open DOI-to-DOI citations) which have cited the retracted publications, along with their distribution according to the *hum_affinity* score of the retracted publication they cite (pie chart).

## 2.3. Annotating the citation characteristics

Once collected the 72 retracted items and their related 678 citing entities, we wanted to characterize such citing entities with respect to their (a) basic metadata and (b) full-text content.

### Gathering citing entities metadata

We retrieved basic metadata via REST APIs from either COCI/MAG, for each citing entity, i.e., DOI (if any), year of publication, title, venue id (ISSN/ISBN), and venue title. Then, using the Retraction Watch database, we annotated whether the citing entity was fully retracted as well.

We also classified the citing entities into areas of study and specific subjects, following the Scimago Journal Classification (https://www.scimagojr.com/), which uses 27 main subject areas (medicine, social sciences, etc.) and 313 subject categories (psychiatry, anatomy, etc.). We searched for the titles and IDs (ISSN/ISBN) of the venues of publication of all the citing entities and classified them into specific subject areas and subject categories. For books/book chapters, we used the ISBNDB service (https://isbndb.com/) to look up the related Library of Congress Classification (LCC, https://www.loc.gov/catdir/cpso/lcco/), and then we mapped the LCC categories into a corresponding Scimago subject area using an established set of rules detailed in (Heibi & Peroni, 2022).

Extracting textual content features

We extracted the abstract of each citing entity and all its in-text citations to the retracted publications in our set, marking the reference pointers to them (i.e., the in-line textual devices, e.g., "[3]" used to refer to bibliographic references), the section where they appear, and their citation context[4]. The citation context is based on the sentence that contains the in-text reference (i.e., the anchor sentence), plus the preceding and following sentences[5]. The definition of this citation context is based on the study of Ritchie et al. (2008). We annotated the first-level sections containing the in-text citation with their type using the categories "introduction", "method", "abstract", "results", "conclusions", "background", and "discussion" listed in (Suppe, 1998) if such section rhetoric was clear by looking at its title, otherwise we used other three residual categories, i.e., "first section", "middle section" and "final section" depending on their position in the citing entity.

---

[4] In case we could not access the full text of a citing entity, e.g., due to paywalls restrictions, the corresponding entity was still considered in our dataset. However, we did not use it for the qualitative post-analysis described in Section "In-text citations" and "Topic modelling". Details about the number of entities for which we could not retrieve are introduced in Section "Citing entities".

[5] Exceptions to this rule, e.g., when the anchor sentence is the last one of a paragraph, are discussed in (Heibi & Peroni, 2022).

Then, we manually annotated each in-text citation with three main features: the citation sentiment conveyed by the citation context, whether the citation context mentioned the retraction of the cited entity, and the citation intent. The annotation of the citation sentiment is inspired by the classification proposed in (Bar-Ilan & Halevi, 2017), and we marked each in-text citation with one of the following values:

- *positive*, when the retracted publication was cited as sharing valid conclusions, and its findings could have been also used in the citing entity;
- *negative*, if the citing entity cited the retracted publication and addressed its findings as inappropriate and/or invalid;
- *neutral*, when the author of the citing entity referred to the retracted publication without including any judgment or opinion regarding its validity.

Then, we annotated with *yes/no* each citing entity if any in-text citation context we gathered from it did/did not explicitly mention the fact that the cited entity was retracted. Finally, we annotated the intent of each in-text citation. The citation intent (or citation function) is defined as the authors' reason for citing a specific publication (e.g., the citing entity *uses a method defined in* the cited entity). For labelling such citation functions we used those specified in the Citation Typing Ontology (CiTO, http://purl.org/spar/cito) (Peroni & Shotton, 2012), an ontology for the characterization of factual and rhetorical bibliographic citations. We used the decision model developed and adopted in (Heibi & Peroni, 2021a) to decide which citation function select to label an in-text citation. Figure 4 shows part of the decision model, it presents the case when the intent of the citation is "Reviewing and eventually giving an opinion on the cited entity" and the citation function is part of one of the following groups: "Consistent with", "Inconsistent with", or "Talking about".

We do not introduce the full details of the labelling process due to space constraints; the complete diagram of the decision model is available at (Heibi, 2022) and an extensive introduction and explanation can be found in (Heibi & Peroni, 2022).

Figure 4. Part of the decision model for the selection of a CiTO (Citation Typing Ontology) citation function for annotating the citation intent of an examined in-text citation based on its citation context. The first large row contains one of the three macro categories ("Reviewing …"); each macro category has a set of subcategories such that each subcategory refers to a set of citation functions. The first row defines what are the citation functions suitable for it through the help of a guiding sentence which needs to be completed according to the chosen sub-category and citation function.

# 3. Results and analysis

We have produced an annotated dataset containing a total of 678 citing entities and 1020 in-text citations to 72 retracted publications. We published a dedicated webpage (https://ivanhb.github.io/ret-analysis-hum-results/) embedding visualizations that enable the readers to view and interact with the results, also available in (Heibi & Peroni, 2021b).

In the next sections, we introduce some important concepts adopted in the description and organization of our results, then we show the results of a quantitative and qualitative analysis of all the data we collected.

## 3.1. Data organization

We defined three periods to distribute the citations to retracted publications:

- period *P-Pre* – from the year of publication of the retracted work to the year before its full retraction (the year of the retraction is not part of this period).
- period *P-Ret* – the year of the full retraction.
- period *P-Post* – from the year after the full retraction to the year of the last citation received by the retracted publication, according to the citation data we gathered.

Each citing entity falls under one of the above three periods. The two periods *P-Pre* and *P-Post* were split into fifths, labelled as "[-1.00, -0.61]", "[-0.60, -0.21]", "[-0.20, 0.20]", "[0.21, 0.60]", and "[0.61, 1.00]". In case the citing entity is part of either *P-Pre* or *P-Post*, then it is also part of a specific fifth, which identifies how close/far that entity is to the events that defining the period.

The division into fifths helped us define a uniform timespan to locate the citing entities independently from the year of retraction of the work they cite and the publication years of the citing and cited entities[6]. For instance, if an entity *A* published in 2011 had cited a retracted publication *R* published in 2002, fully retracted in 2012, then *A* is part of the last fifth (i.e., "[0.61, 1.00]".) of *P-Pre*. This means that *A* has cited *R* in the last fifth, immediately before the formal retraction of *R*.

## 3.2. Descriptive statistics

We have classified the distribution of the citing entities in the three periods (i.e. *P-Pre, P-Ret, and P-Post*) as a function of the humanities disciplines used in Retraction Watch, as shown in Figure 5. *Religion* was the discipline that received the highest number of citations (375), while *history* had the highest number of retracted items (20).

In Figure 6 we have classified the entities citing a retracted publication in each discipline according to their subject areas. *Arts and humanities* and *Social Sciences* (AH&SS) were highly represented in both the *P-Pre* and *P-Post* periods of almost all the retracted publications' disciplines. However, we noticed some exceptions to this rule in *P-Pre* in *Journalism* (10% of citing entities were AH&SS publications)*, P-Post* in *Arts* (13% AH&SS publications)*,* and *P-Pre* and *P-Post* of *Architecture* (no AH&SS publications in both periods).

Since we expected, as also highlighted in previous studies (e.g., Ngah & Goi,1997), that a good part of the citations to humanities publications come from AH&SS publications, we decided to deepen into the obtained results before moving on the next stage. As shown in Figure 5, we noticed that *Journalism* has a completely different behavior compared to the other disciplines. Indeed, the citations of *Journalism* have cited three retracted publications: two with a *hum_affinity* of 3, and one with a *hum_affinity* of 2. The latter article was *"Personality, stress and disease: description and validation of a new inventory"* (Grossart-Maticek & Eysenck, 1990). This article has 130 citations (almost 95% of all the citations in *Journalism*). Retraction Watch has labeled this article with the additional two subject areas: *Public Health and Safety*, and *Sociology*, therefore *Journalism* represents the only humanities subject. A further investigation in the full text of the paper revealed the fact that this article is highly related to health sciences, and *Journalism* has a marginal (almost absent) relevance in it. Considering these discovered facts, we felt that this article could represent a significant bias to our analysis. Therefore, to limit its impact on the results we decided to exclude it from our analysis.

---

[6] A detailed explanation regarding the calculation of the periods is discussed in (Heibi & Peroni, 2022).

As a further check, we have investigated all the retracted publications of all the humanities disciplines in Figure 6 having citations from *Arts and Humanities* publication less than 20% in either *P-Pre* or *P-Post*. *Arts* and *Architecture* are the two disciplines falling in this category. After a manual check, we detected the article *"A systematic review on post-implementation evaluation models of enterprise architecture artefacts"* (Nikpay et al., 2020), classified under *Architecture,* yet while reading its full text we found little evidence supporting the proposed labelling, since it was a computer science study. Therefore, we decided to also exclude this article from our analysis.

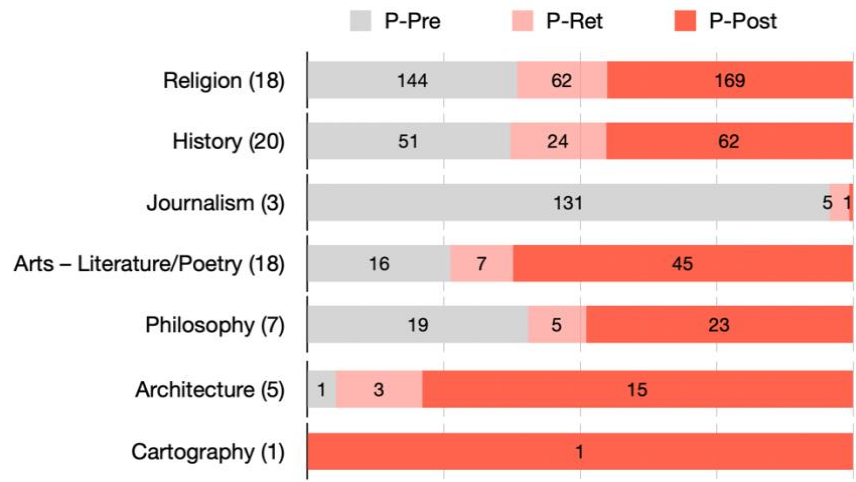

**Figure 5.** The number of citing entities in *P-Pre* (before the year of retraction), *P-Ret* (in the year of retraction), and *P-Post* (after the year of retraction) for each different humanities discipline specified to the retracted publication as gathered from Retraction Watch.

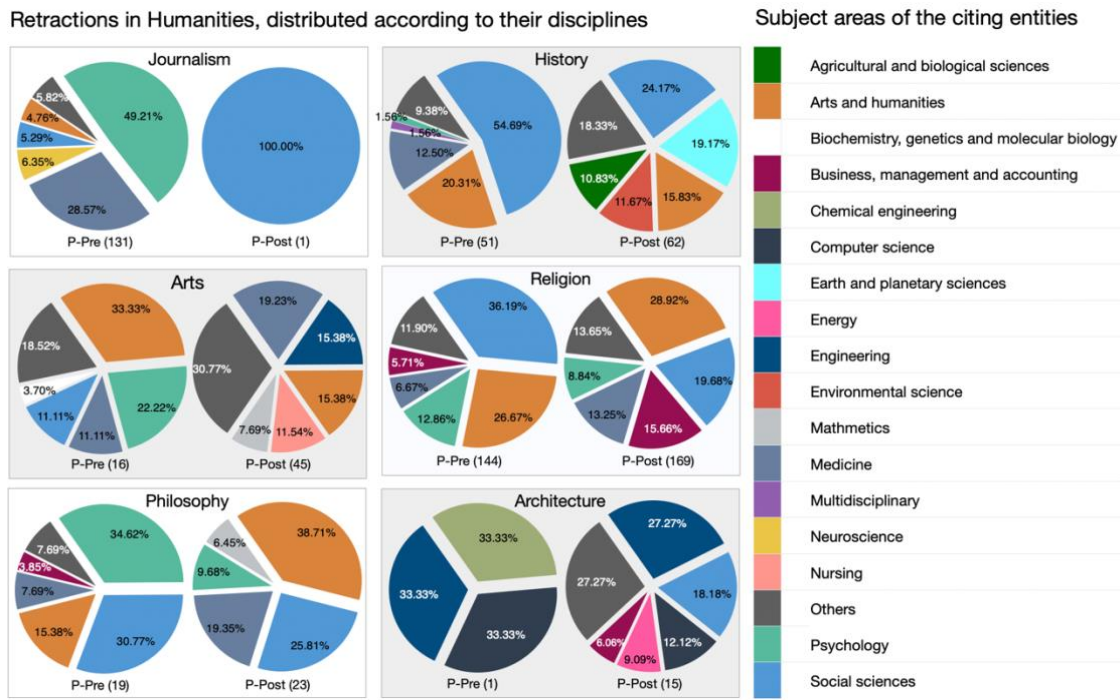

**Figure 6.** The subject areas distribution of the citing entities of the retracted publications in *P-Pre* (before the year of retraction) and *P-Post* (after the year of retraction) for each different humanities discipline as specified in Retraction Watch. The number of citing entities is mentioned between brackets.

After this data refinement, our final data have been reduced to a total of 546 citing entities and 786 in-text citations to 70 retracted publications. Considering the final data and the classification of the retracted publications based on their humanities discipline, we investigated another aspect; in Figure 7 we have plotted the total number of citations gained by each humanities discipline as a function of the number of years passed after the date of retraction. This trend is compared to the average time of retraction for each humanities discipline. From Figure 7, we noticed that on average disciplines such as religion and philosophy reported their peak in the year before their retraction, while this trend is the opposite for history, arts, and architecture.

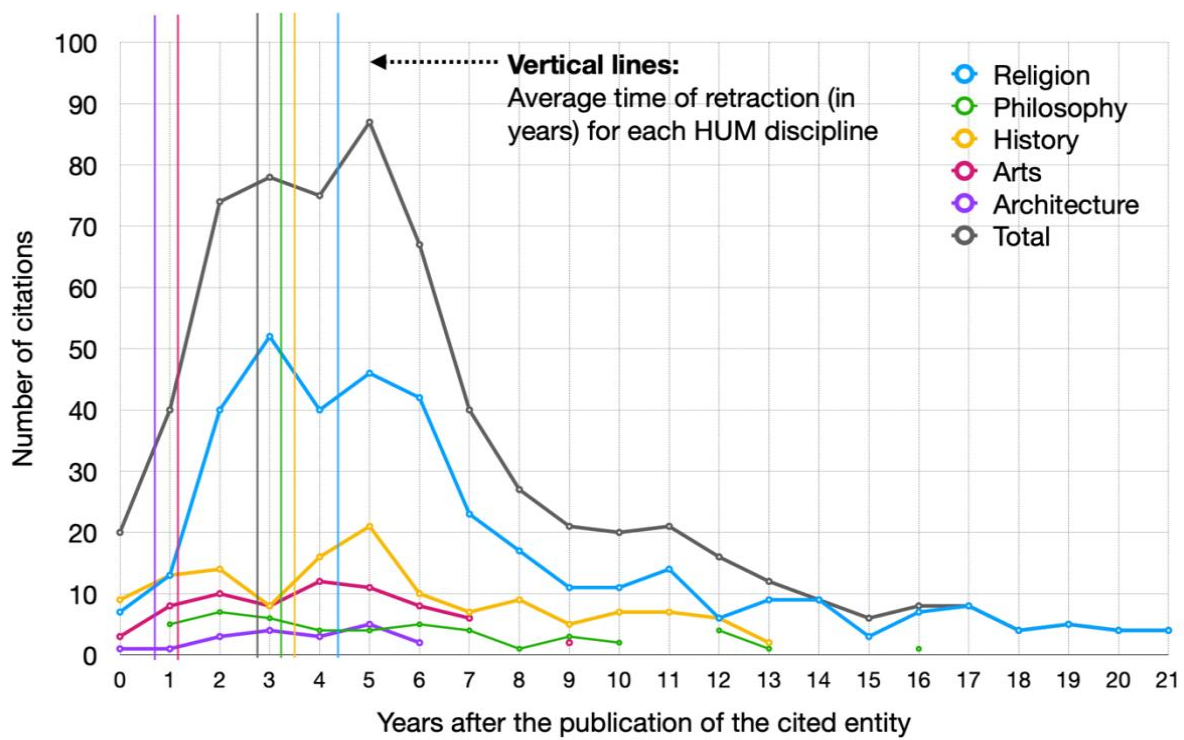

**Figure 7.** The total number of citations gained by the retracted publications, grouped according to their humanities discipline (represented with different colors), as a function of the number of years passed after their date of retraction. The vertical dotted lines represent

the average time of retraction of each humanities discipline. The gray line sums up all the humanities disciplines together.

To infer other interesting statistics regarding the obtained results, we treated the citing entities and the in-text citations they contain as two different classes, and we present descriptive statistics of these two classes in the following sub-sections.

### Citing entities

We examined the distribution of the citing entities to retracted publications as a function of two features: (1) the periods (i.e., *P-Pre*, *P-Ret*, and *P-Post*), further classified into those that mentioned the retraction or for which we could not access their full-text, and (2) their subject areas. The results are shown in Figure 8.

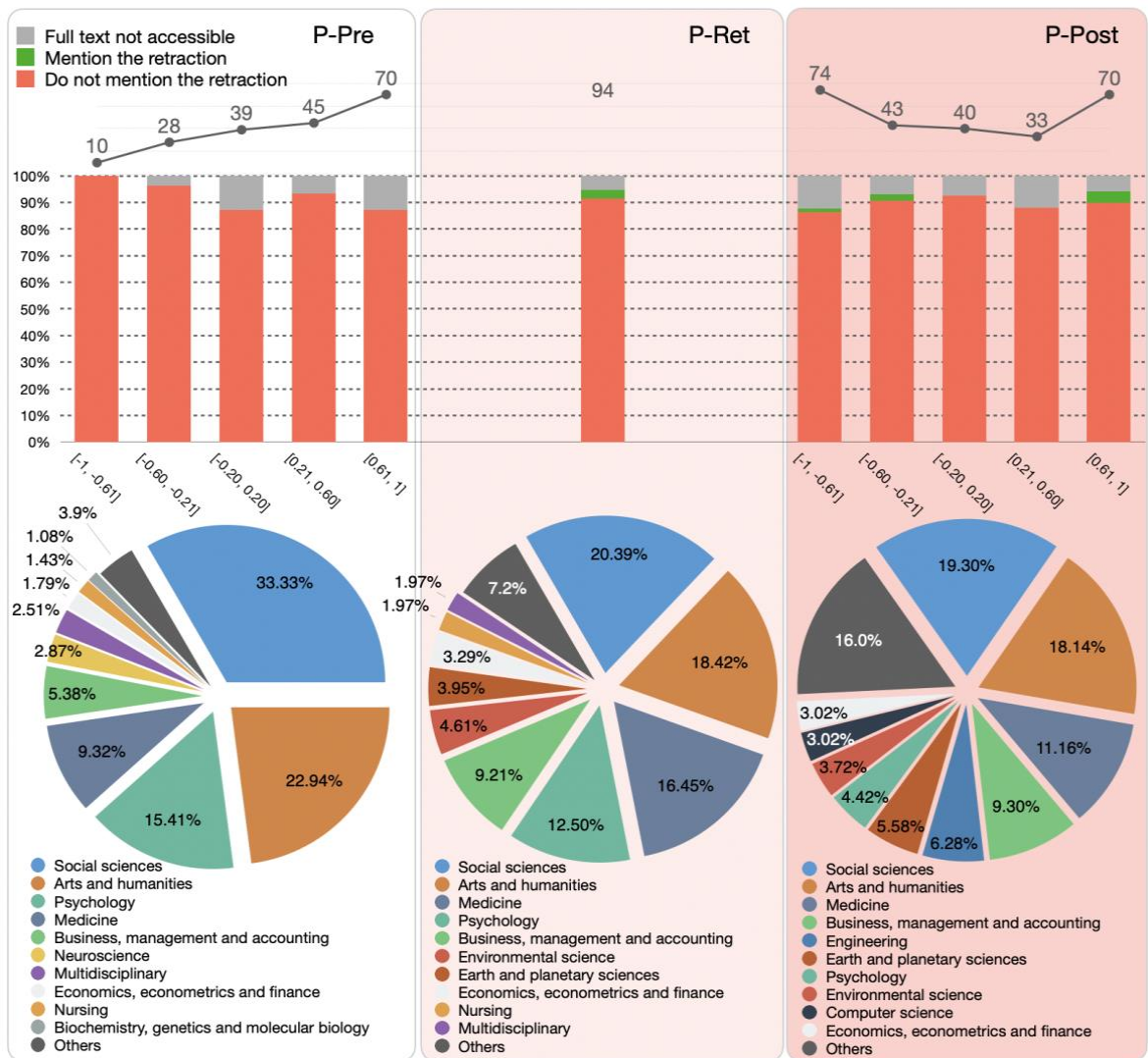

**Figure 8.** A descriptive statistical summary of the distribution of the citing entities to retracted publications in the three periods (*P-Pre*, *P-Ret*, and *P-Post,* i.e., before/during/after the year of retraction), also considering their subject areas. The bar charts on top highlight the citing entities that either did/did not mention the retraction and those which we could not retrieve the full text.

The number of the citing entities before the retraction (192, period *P-Pre*) was lower than the number of the citing entities after the retraction (260, period *P-Post*). Along *P-Pre* and *P-Ret*, we noticed a continuous increment in the overall number of citing entities, that suddenly started decreasing after the first fifth of *P-Post*, yet the numbers were in line with the ones observed in the third and fourth fifth of *P-Pre*. The last fifth of *P-Post* is an exception to the declining trend, with an unexpected high pick. This result was due to the fact that 27 retracted items received only one citation in *P-Post* and, in these cases, that citation always represented the last citation received, which is the final border of *P-Post*.

The full text of 8.42% of the citing entities was not accessible. For those that have successfully retrieved the full-text, our results showed that a relatively low percentage mentioned the retraction of the cited entity – 2.25% of the total number of citing entities in *P-Ret* and *P-Post*.

Looking at their subject areas, we noticed that the citing entities started to spread into a higher number of subject areas (i.e., additional 9) in *P-Post* compared to *P-Pre*, where the residual category *Others* contained 16% of the citing entities. The *Arts and Humanities* subject area had a similar percentage throughout all the three periods (22.94%, 18.42% and 18.14%), and it represents, together with *Social Sciences*, the 2 most representative subject areas in *P-Ret* and *P-Post*. We also noticed an important drop-down in *Psychology*, from 15.41% in *P-Pre* to a 4.42% in *P-Post*.

### In-text citations

We focused on the distribution of the in-text citations as a function of three features: (1) the periods (i.e., *P-Pre*, *P-Ret*, and *P-Post*), (2) the citation intent, and (3) the section containing the in-text citation. The results of the three distributions have been further classified according to the in-text citation sentiment (i.e., *negative/neutral/positive*), as shown in Figure 9.

The overall trend in the number of in-text citations along the three periods was close to the one we observed for the citing entities (shown in the previous section), although the differences between *P-Pre* and *P-Post* were even more marked. As introduced in the previous section, the pick in the last fifth of *P-Post* was due to the retracted items receiving only one citation in *P-Post*. Even though the overall percentage of negative citations was low, it had a higher presence in *P-Pre* (4.5%). Generally, most in-text citations were tagged as *neutral*, and very few were *positive* (0.75%).

The citation Intents "*obtains background from*" and "*cites for information*" were the two most dominant ones in the three periods, and they respectively represented 31.29% and 22.64% of the total number of in-text citations, respectively. The citation intent "*cites for information*" increased its presence moving from 17.8% in *P-Pre* to 27.20% in *P-Post*.

Considering the citation sections, we can clearly see that the in-text citations were mostly located in the "*Introduction*" section in all the three periods. The in-text citations in the section "*Introduction*" decreased a lot after *P-Ret* moving from 30.15% in *P-Pre* to 22.13% in *P-Post*. Instead, the in-text citations contained in the section "*Discussion*" have an increasing trend, from 6.87% in *P-Pre* to 15.20% in *P-Post*.

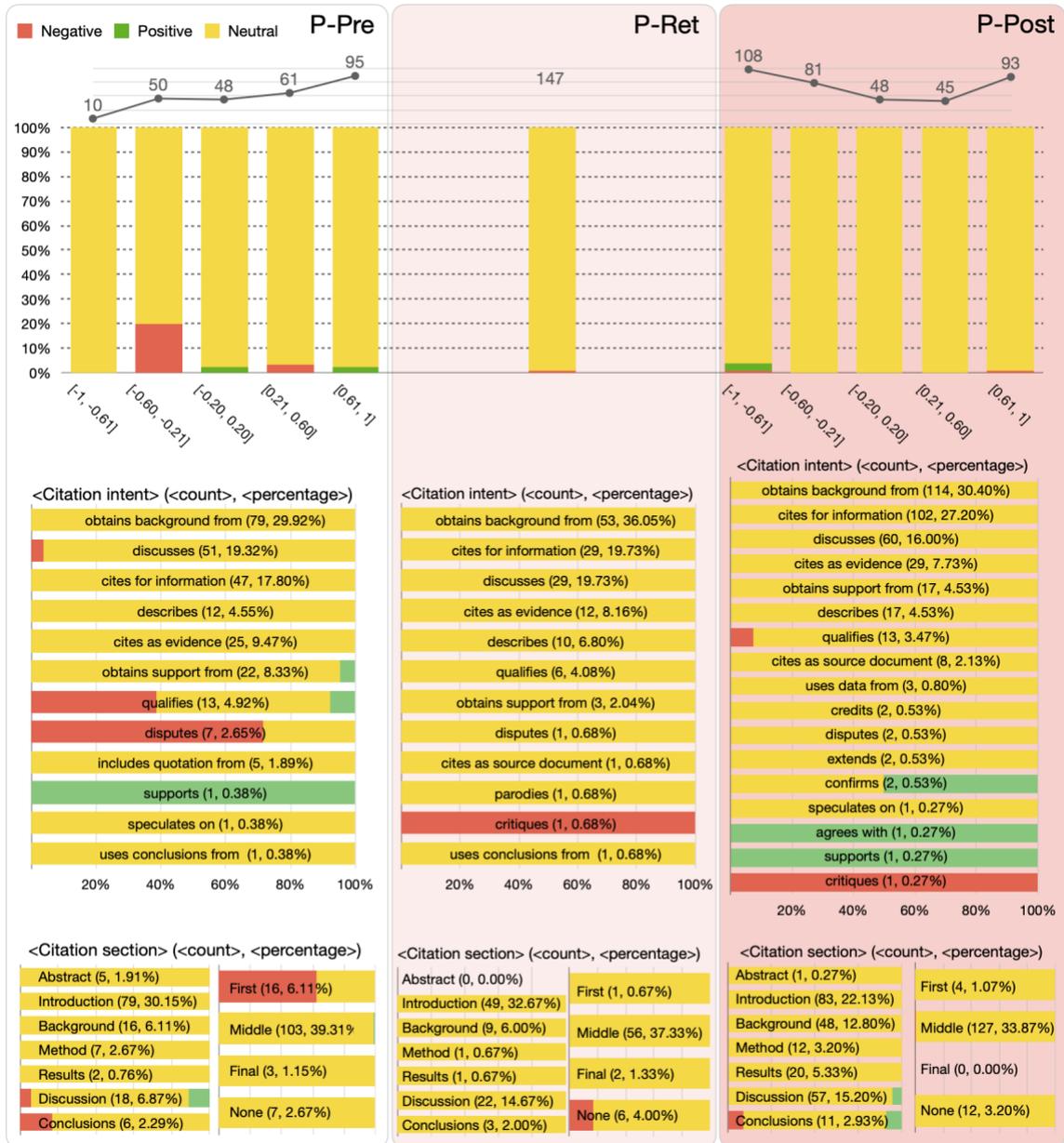

**Figure 9.** A descriptive statistical summary for the distribution of the in-text citations contained in the citing entities to the retracted publications in the three periods (*P-Pre*, *P-Ret*, and *P-Post,* i.e., before/during/after the year of retraction), according to their intent, and section. The sentiment of the in-text citations is also highlighted.

## 3.3. Topic models of citing entities' abstracts and their citation contexts

A topic modeling analysis is a statistical modeling approach for automatically discovering the topics (represented as a set of words) that occur in a collection of documents. We used it with our data to understand what the evolution of the topics in time was and whether it was dependent, in some way, to the retraction received by the publications considered.

A standard workflow for building a topic model is based on three main steps: tokenization, vectorization, and topic model I creation. The topic model we have built is based on the Latent Dirichlet Allocation (LDA) model (Jelodar et al., 2019). In the tokenization process we have converted the text into a list of words, by removing punctuations, unnecessary characters, and stop words, and we also decided to lemmatize and stem the extracted tokens. In the second step, we created vectors for each of the generated tokens using a Bag of Words (BoW) model [7], which we considered appropriate to model our study considering our direct experience in previous findings (Heibi & Peroni, 2021a) and the suggestions by Bengfort et al. (2018) on the same issue. Finally, to build the LDA topic model, we determined in advance the number of topics to retrieve according to the examined corpus using a popular method based on the value of the topic coherence score, as suggested in (Schmiedel et al., 2019), which can be used to measure the degree of the semantic similarity between high-scoring words in the topic.

---

[7] Brownlee, J. (2019). *A Gentle Introduction to the Bag-of-Words Model*. 30. https://machinelearningmastery.com/gentle-introduction-bag-words-model/

We built and executed two LDA topic models, one using the abstracts of the entities citing the retracted publications (with 16 topics) – named *TM-Abs,* and another using the citation contexts where the in-text reference pointers to retracted publications were contained (with 20 topics) – named *TM-Cits*. For creating the topic models, we used MITAO (Ferri et al., 2020) ([https://github.com/catarsi/mitao](https://github.com/catarsi/mitao)), a visual interface to create a customizable visual workflow for text analysis. With MITAO we have generated two visualizations, i.e., LDAvis (Latent Dirichlet Allocation Visualization) (Sievert & Shirley, 2014), for an overview of the topic modeling results, and MTMvis (Metadata-based Topic Modeling Visualization), for a dynamic and interactive visualization of the topics based on customizable metadata.

### Citing entities abstracts

The total number of available abstracts in our dataset was 509. We extended the list of MITAO's default English stop-words (e.g., "the", "is", etc.) with ad-hoc stop-words devised for our study such as "method", "results", "conclusions", etc., which represents the typical words that might be part of a structured abstract.

Figure 10 shows the topics distribution represented in the two-dimensional space of LDAvis. Using LDAvis interface, we set the parameter $\lambda$ to 0.3 to determine the weight given to the probability of a term under a specific topic relative to its lift (Sievert & Shirley, 2014), and retrieved the 30 most relevant terms of each topic. We gave an interpretation and a title to each topic by analyzing its related terms, that we avoid introducing here due to space limitation, but they are available in (Heibi & Peroni, 2021b). Topic 6 ("Leadership organization, and management") was the dominant topic. The topics were distributed in four main clusters, as shown in Figure 10:

- one composed by topics 2 ("Socio-political issues related to leadership") and 6, concerned issues related with leadership, work organization, and management form a socio-political point of view;
- a large one composed by topics 1("Socio-political issues possibly related to Vietnam"), 4 ("History of the Jewish culture"), 5 ("Music and psychological diseases"), 11("Family and religion"), etc. Treat several subjects from different domains close to social sciences, political sciences and psychology;
- other two clusters composed by one topic each, i.e. topic 16 ("Geography and climatic issues") and topic 3 ("Colonial history").

Figure 11 shows the chart generated using MTMvis. We plotted the topics distribution as a function of the three periods. At a first analysis, we noticed how topics 6 and 16 incremented their distribution along the three periods. On the other hand, topics 1 and 11 decreased their percentage throughout the three periods.

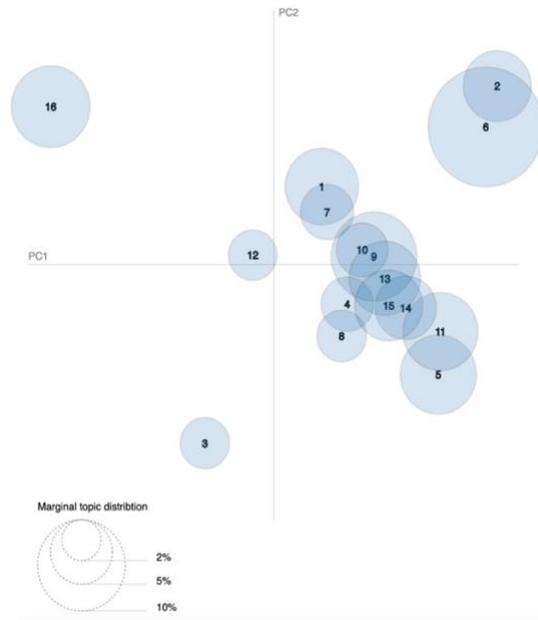

**Figure 10.** The 16 topics of *TM-Abs* (LDA topic modeling on the abstracts of the citing entities). The visualization is taken from LDAvis, and it shows the topic distribution in a two-dimensional space.

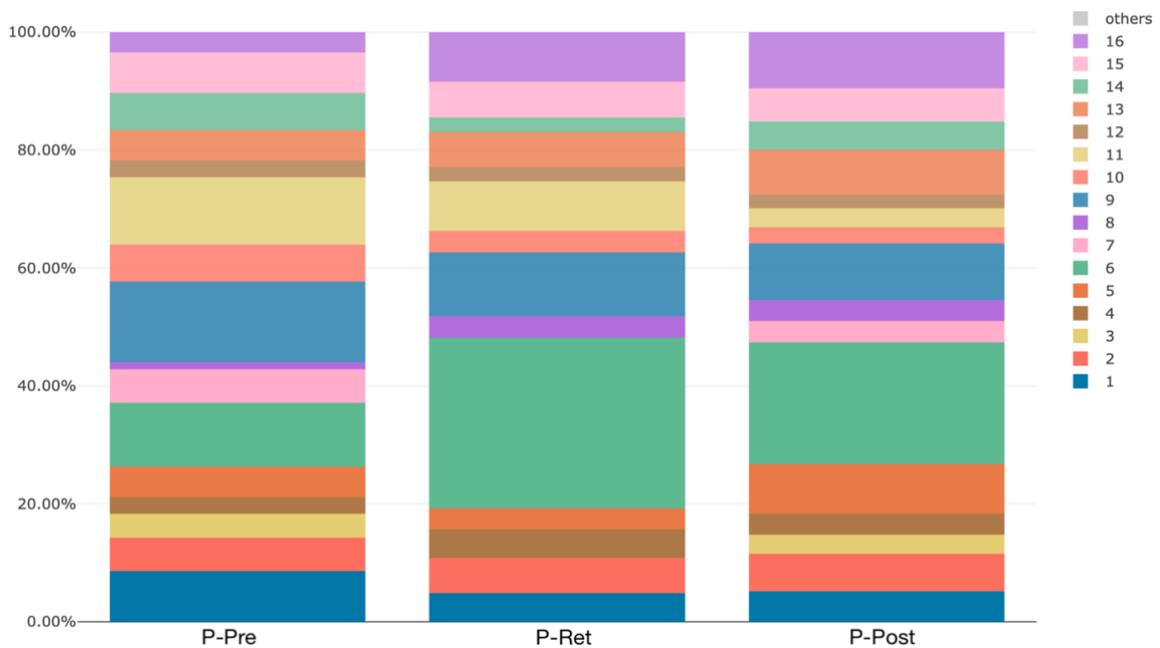

**Figure 11.** The MTMvis chart created over the 16 topics of *TM-Abs* (LDA topic modeling on the abstracts of the citing entities). The topics are plotted as a function of the three periods (represented on the x-axis)

In-text citation contexts

The total number of in-text citation contexts in our dataset, we used as input to produce the second topic model, was 786. As we did with the abstracts, we have defined and used a list of ad-hoc stop-words, which included all the given and family names of the authors of the cited publications.

Figure 12 shows the topics represented in the two-dimensional space of LDAvis. As we did for the abstracts' topic modeling, we set $\lambda$ to 0.3 and interpreted each topic by analyzing its 30 most relevant terms (Heibi & Peroni, 2021). In this case, we noticed that the topics are less overlapping and more distributed along all the axis of the visualization. Topic 12 ("Leadership organization, and management") is the most representative (11.7%) and was very distant from the other topics. The bottom right part of the graphics – with topics 2 ("Countries in conflict"), 15 ("War and terrorism"), 17 ("War and history"), 18 ("History of Europe"), 20 ("War and army conflicts") – are mostly close to the history studies, especially discussion toward army conflicts. The part on the top of the graphics contains several single-topic clusters, such as topic 5 ("Gender social issues") and 9 ("Geography and climatic issues").

Figure 13 shows the chart generated using MTMvis, where we plotted the topic distribution as a function of the three periods. We noticed a continuous decrement in topics 7 ("Family and religion") and 18 along the three periods. Topic 3 ("Drugs/Alcohol and psychological diseases") had a high decrement right after P-Ret. On the other hand, we noticed an increment in topics 5, 9, and 11 ("Music and psychological diseases") – although the latter topic had a higher percentage in P-Ret than in P-Post.

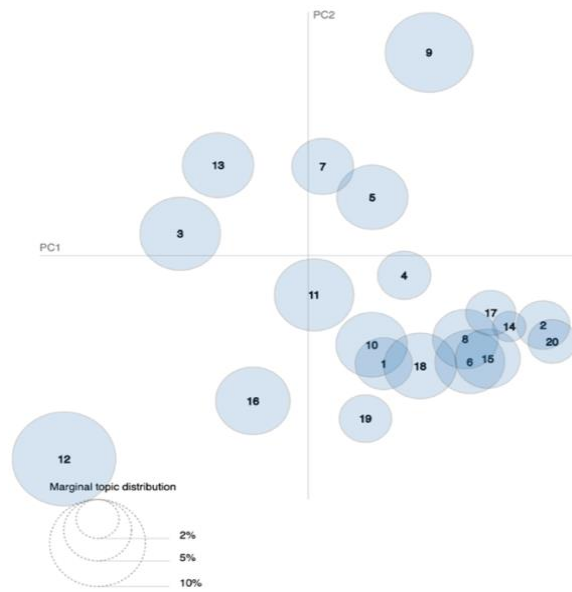

**Figure 12.** The 20 topics of *TM-Cits* (LDA topic modeling on the in-text citation contexts). The visualization is taken from LDAvis, it shows the topic distribution in a two-dimensional space.

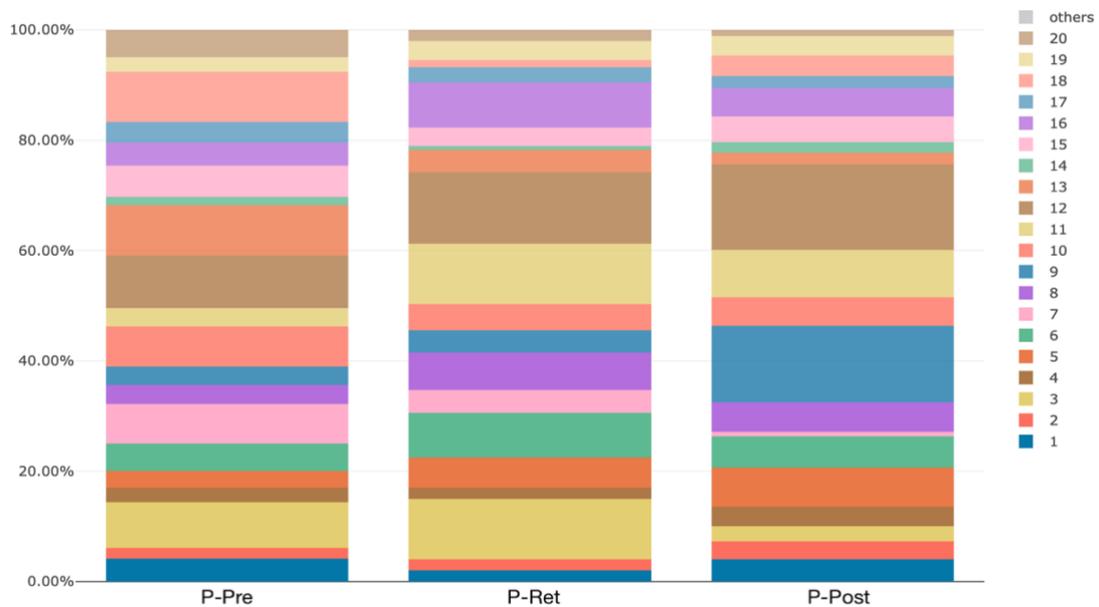

**Figure 13.** The MTMvis chart created over the 20 topics of *TM-Cits* (LDA topic modeling on the in-text citation contexts). The topics are plotted as a function of the three periods (represented on the x-axis).

# 4. Discussion and conclusion

In this section, we address separately each of our research questions RQ1-RQ3 presented in Section "Introduction". We conclude the section by discussing the limits of our work and by sketching out some future works that might help us overcome these issues.

## 4.1. Answering RQ1: citing retracted publications in the humanities

It seems that, on average, retracted publications in the humanities did not have a drop of citations after their retraction (Figure 8) and only 2.25% of the citing entities – 5 *arts and humanities* publications and 3 related to health sciences subject areas (e.g., *medicine, psychology, nursing*, etc.) – mentioned the retraction in the citation context. In addition, we noticed that the negative perception of a retracted work, although limited in the data we have, happened before their retraction if the cited entity had a low affinity to the humanities domain. The fact that we reported few negative citations in *P-Post* is along the line of other studies (Bordignon, 2020; Schneider et al, 2020; Luwel et al., 2019).

Citing entities talking about retraction usually *discussed* the cited entity rather than *obtaining background* material from it or generic *informative* claims (Figure16). Most of the in-text citations marked as *discusses* occurred in the *Discussion* section (as shown in Figure 15), and from *TM-Cits* we noticed the emerging of topic 6 ("The retraction phenomenon") in *Discussion* sections only in *P-Post* – in other words, the retraction was not mentioned in the *Discussion* section before the retraction, and the retraction event might have been the trigger of a higher discussion from the citing entities.

From the distribution of the subject areas of the citing entities over the three periods (Figure 8), we noticed that *social sciences* and *arts and humanities* had almost the same percentages in the *P-Ret* and *P-Post* periods, which is less than their percentages in *P-Pre*, suggesting that the retraction event did have an impact on these subject areas. However, other subject areas such as *psychology* decreased in *P-Ret* and more in *P-Post*, that may be an indicator of a higher concern by these subject areas toward the citation of retracted publications. This is evidenced by the observation of the *TM-Abs* topics distribution for the citing entities assigned to *psychology* (Figure 14), with a clear decrement in the topics related to health sciences such as topics 10 and 11, while others such as topics 6 and 9 (close to socio-historical discussions with no relation to health sciences) increased their presence in *P-Ret* and *P-Post*. In other words, not only the overall number of the citing entities from the health sciences domain decreased after the retraction, but their subject areas moved from the health sciences domain to subjects that are closer to the *social sciences* and *arts and humanities* domain.

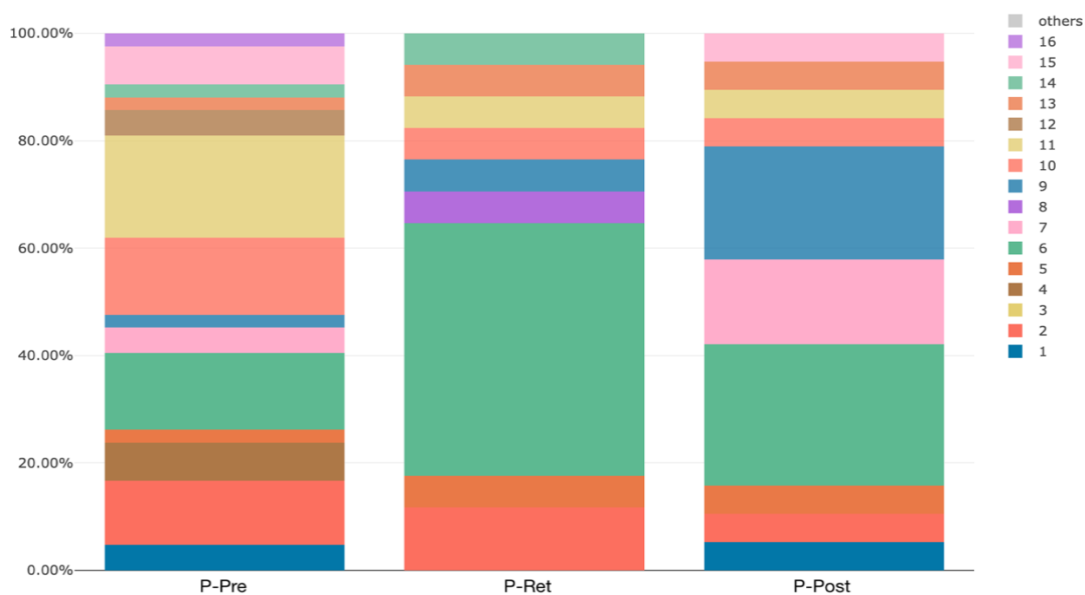

**Figure 14.** A filtered MTMvis to show the distribution of the topics of *TM-Abs* (LDA topic modeling on the abstracts of the citing entities) as a function of the three periods. The visualization is built considering only the documents (i.e., abstracts) that have *Psychology* as subject areas.

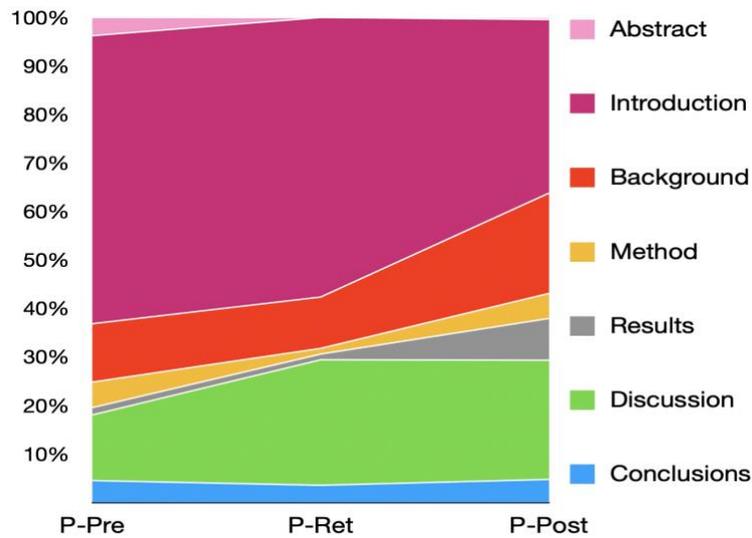

**Figure 15.** The distribution of the main (positional sections are not included, e.g., *first section*) in-text citation sections over the three periods. The percentages of in-text citations having a corresponding annotated main section for each period (i.e., *P-Pre*, *P-Ret* and *P-Post*) are respectively 50.76%, 56.68%, and 61.86%.

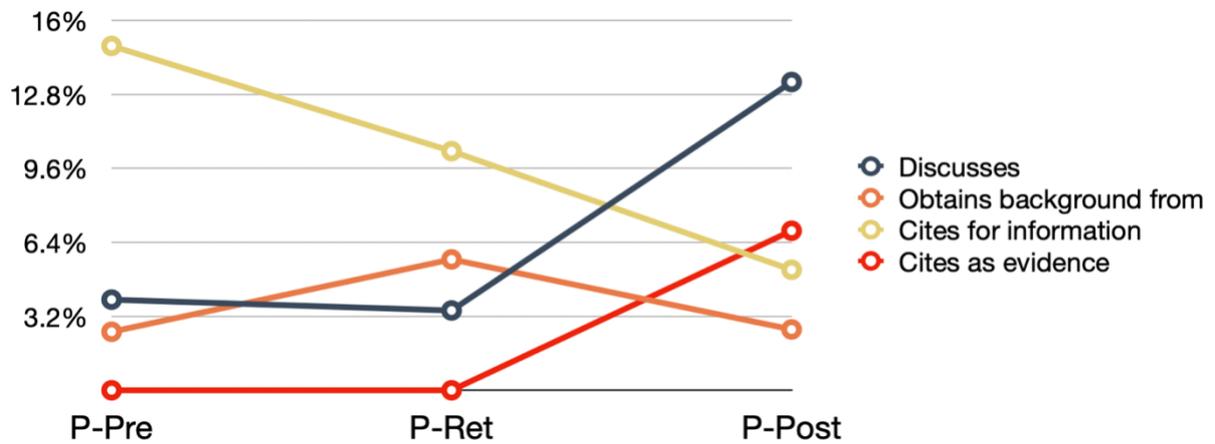

**Figure 16.** The distribution of topic 6 ("The retraction phenomenon") of *TM-Cits* (LDA topic modeling on the in-text citation contexts) over the three periods for the 4 citation intents which have been used the most.

## 4.2. Answering RQ2: citation behaviors in the humanities

As shown in Figure 6, *religion* and *history* had a very similar distribution pattern. In both, the citing entities belonging to *social sciences* had an important decrement in *P-Post*, and for that period the *TM-Cits* of these entities does not include topic 3 ("Drugs/Alcohol psychological diseases") for *religion* and topic 7 ("Family and religion") for *history*. We can speculate that *social sciences* studies significantly reduced its percentage due to a higher concern toward sensitive social subjects such as health care, family, and religion.

*Arts* had the highest number of citations in *P-Post*, although we reported an important drop in the *arts and humanities* citing entities, in favor of subject areas such as *medicine, nursing* and *engineering* (Figure 6). On the other hand, for *philosophy,* we had a completely different situation: citing entities labeled as *arts and humanities* incremented a lot in *P-Post* at the expense of citing entities from *psychology*. For the *arts* discipline, topic 11 ("Music and psychological diseases") of *TM-Cits* is the reason for the positive trend of *P-Post*. In other words, arts (and especially music) had been discussed with relation to psychological and medical diseases.

In Figure 17, we show the distribution of topic 6 ("The retraction phenomenon") as a function of the three periods and considering the four humanities disciplines with the higher number of citing entities. Topic 6 increased a lot in *P-Post* in *philosophy*, in *religion* it had a steady trend, while in *history* and *arts* had the pick in *P-Ret* and had a lower, yet relatively high, percentage in *P-Post*. These results might suggest that the entities which cite retracted publications in *philosophy*, *arts*, and *history*, (which following the results of the topic modeling analysis produced topics close to STEM disciplines) were those showing major concerns toward the retraction – in the case of *history* and *arts*, starting from the year of the retraction.

Considering these hypothesis, we can interpret the fact that *history* and *arts* reached their pick of citations after their year of retraction (Figure 7), as a sign of awareness/acknowledgment regarding the retraction rather than an unconsciousness usage of the retracted publications, at least for part of these citations.

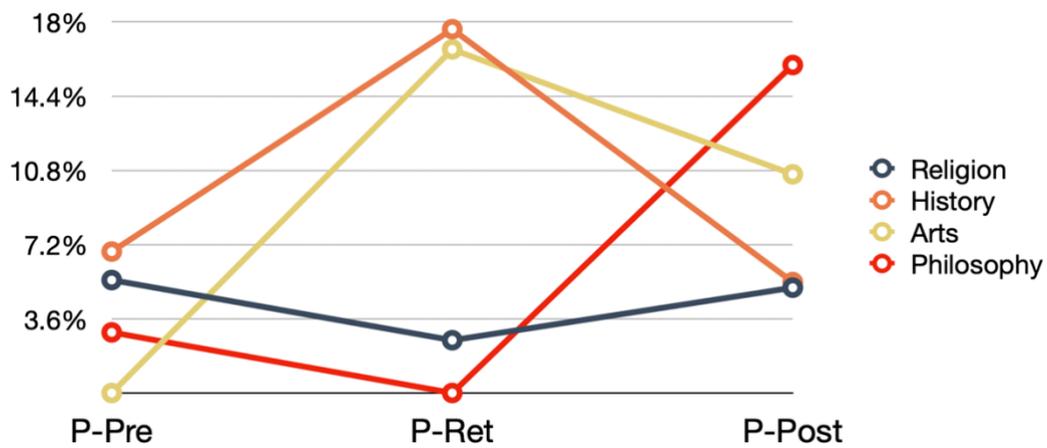

**Figure 17.** The distribution of topic 6 ("The retraction phenomenon") of *TM-Cits* (LDA topic modeling on the in-text citation contexts) over the three periods for the humanities disciplines: Religion, History, Arts and Philosophy.

## 4.3. Answering RQ3: comparing STEM and the humanities

Our findings showed that the retraction of humanities publications did not have a negative impact on the citation trend (Figure 8). The opposite trend was observed in other disciplines, according to prior studies, such as biomedicine (Dinh et al., 2019) and psychology (Yang et al., 2020). However, prior studies such as (Heibi & Peroni, 2021a) and (Schneider et al., 2020), also observed that in the health sciences domain were cases where either a single or a few popular cases of retraction were characterized by an increment of citations after the retraction. This might suggest that the discipline related to the retracted publication is not the only central factor to consider for predicting the citation trend after the retraction. Other factors might play a crucial role, such as the popularity and the media attention to the retraction case, as it has been discussed in the studies by Mott et al. (2019) and Bar-Ilan and Halevi (2017).

The work by Bar-Ilan and Halevi (2018) analyzed the citations of 995 retracted publications and found the same growing trend in the citations in the post retraction period. However, they did not analyze the retraction according to different and separate disciplines. As such, we might consider such results as a representation of a general trend of retracted publications, that confirms the general observations we derived from our data. In addition, considering the results we have obtained for the specific humanities disciplines, it seems like the potential threats and damage of retracted materials has been perceived more seriously by others (i.e., citing entities) when the retracted publications have been linked to a sensitive area of study and to STEM domain. This final observation remarks the different behaviors that might occur when a retracted publication manifests a higher relation to STEM.

## 4.4. Limitations and future developments

There are, indeed, some limitations in our studies that may have introduced some biases. First, compared to other fields of study, bibliographic metadata in the humanities have a limited coverage in well-known citation databases (Hammarfelt, 2016). This fact led to some limitations when applying a citation analysis in the humanities domain (Archambault & Larivière, 2010). With this regard, a coverage analysis and comparison of the citations in the humanities domain in COCI and MAG might be highly valuable. Other data sources such as OpenAlex (Priem et al., 2022) – a free and open catalog of the world's scholarly papers, researchers, journals, and institutions, could be considered. Pragmatically, for what our study was concerned, we surely collected fewer citing entities than those that had in fact cited the retracted publications. In addition, we have considered only open citation data, therefore the citation coverage could significantly improve with the addition of non-open citation data. The availability of a larger amount of data could have strengthened and improved the quality of our results.

The selection of the retracted publications was another crucial issue since we faced two major problems: (1) some inconsistencies in the data provided by Retraction Watch, and (2) the presence of retracted publications labeled as humanities that, at a close analysis, actually belonged to a different discipline. The first descriptive statistical results, our manual check, and the definition of the humanities affinity score, helped us limit the biases of these two issues. However, we could improve the approach adopted by using additional services such as Elsevier's ScienceDirect – as done in (Bar-Ilan & Halevi, 2018) – and to increase the threshold of the humanities affinity level to exclude border cases

A citation analysis toward the retraction in the humanities domain is something that have been rarely discussed in the past, therefore the discussion of our results included a comparison with similar works which considered also different domains or retraction cases. Such works have not addressed the humanities domain or were based either on a single or a limited set of retraction cases. Works which considered other domains did not include most of the features that we have analyzed in this work. e.g., the citation intent, which made the comparison with them difficult. We aim that this study and others to be done in this field can favor a comparison and improvement in the understanding of the retraction phenomenon in the humanities domain.